\documentstyle[12pt,aasms4] {article}

\begin{document}

\title{Faint Infrared Flares from the Microquasar GRS
1915+105}

\author{S.S. Eikenberry\altaffilmark{1}, K. Matthews\altaffilmark{2}, M. Muno\altaffilmark{3}, P.R. Blanco\altaffilmark{4}, E.H. Morgan\altaffilmark{3}, R.A. Remillard\altaffilmark{3}}

\altaffiltext{1}{Astronomy Department, Cornell University, Ithaca, NY  14853}
\altaffiltext{2}{Physics Department, California Institute of Technology, Pasadena, CA  91125}
\altaffiltext{3}{Physics Department, Massachusetts Institute of Technology, Cambridge, MA  02139}
\altaffiltext{4}{CASS, UC San Diego, La Jolla, CA 92093}

\begin{abstract}

	We present simultaneous infrared and X-ray observations of the
Galactic microquasar GRS 1915+105 using the Palomar 5-m telescope and
Rossi X-ray Timing Explorer on July 10, 1998 UT.  Over the course of 5
hours, we observed 6 faint infrared (IR) flares with peak amplitudes
of $\sim 0.3-0.6 $ mJy and durations of $\sim 500-600 $ seconds.
These flares are associated with X-ray soft-dip/soft-flare cycles, as
opposed to the brighter IR flares associated with X-ray
hard-dip/soft-flare cycles seen in August 1997 by Eikenberry et
al. (1998).  Interestingly, the IR flares begin {\it before} the X-ray
oscillations, implying an ``outside-in'' origin of the IR/X-ray cycle.
We also show that the quasi-steady IR excess in August 1997 is
due to the pile-up of similar faint flares.  We discuss the
implications of this flaring behavior for understanding jet formation
in microquasars.

\end{abstract}


\section{Introduction}

	As the archetypal Galactic microquasar, GRS 1915+105 offers
unique observational opportunities for investigating the formation of
relativistic jets in black hole systems.  To date, two types of
ejection events have been observed from this system.  The first of
these, the ``major'' ejections, produce bright ($\sim 1$ Jy)
resolvable radio jets which move with apparent velocities of $v_{\rm
app} = 1.25 c$ and actual space velocities of $v \sim 0.9 c$ (Mirabel
\& Rodriguez, 1994; Fender et al., 1999).  The jets transition quickly
from optically thick to optically thin spectra and then fade on
timescales of several days.  Due to the rarity of these events,
coordinated pointed X-ray observations have not been possible to date.

	The second type of ejection event consists of X-ray
oscillations with hard power-law dips and thermal flares, and
associated synchrotron flares in the infrared (Eikenberry et al.,
1998a,b) and radio bands (Mirabel et al., 1998; Fender \& Pooley,
1998).  We refer to these events as ``Class B'' flares to distinguish
them from the larger ``Class A'' major ejection events.  These smaller
events have peak intensities in the range $\sim 100-200$ mJy from the
infrared (IR) to radio bands, and the time of peak flux exhibits
apparent delays as a function of wavelength which may indicate the
expansion of a synchrotron bubble (Mirabel et al., 1998).  The flares
fade on timescales of several minutes and tend to repeat on timescales
from $\sim 30-50$ minutes (i.e. Pooley \& Fender, 1997; Eikenberry et
al., 1998a).

	In this paper, we present a third type of IR flare from GRS
1915+105 -- faint (sub-milliJansky) IR flares associated with X-ray
soft-dip/soft-flare cycles.  In Section 2, we present the observations
and analysis of these flares.  In Section 3, we discuss the
implications of the flares for understanding relativistic jet
formation in microquasars.  In Section 4, we present our conclusions.

\section{Observations and Analysis}

\subsection{July 1998 Observations}

	We observed GRS 1915+105 on the nights of 8-12 July 1998 UTC
using the Palomar Observatory 5-m telescope and the Cassegrain
near-infrared array camera in the K ($2.2 \mu$m) band.  Details of
these observations and the data reduction will be presented in
Eikenberry et al. (2000), and we summarize them here.  We configured
the camera to take 128x128-pixel (16x16-arcsec) images at a rate of 1
frame per second, with absolute timing provided by a WWV-B receiver
with $\sim 1$ ms accuracy.  We observed GRS 1915+105 in this mode for
approximately 5 hours each night, obtaining $\sim 1.5 \times 10^4$
frames per night.  The field of view was large enough to capture both
GRS 1915+105 and several nearby field stars, including ``Star A'',
which has a magnitude of $K = 13.3$ mag (Eikenberry \& Fazio, 1997;
Fender et al., 1997).  After standard processing (sky/bias
subtraction, flat-fielding, interpolation over bad pixels and cosmic
ray hits) we used the nearby stars to perform differential photometry
on GRS 1915+105, with the overall absolute calibration provded by Star
A.  We present the resulting flux density for GRS 1915+105 on July 10,
1998 UTC with 10-second time-resolution in Figure 1(a).  We obtained
X-ray observations on the same nights using the PCA instrument on the
Rossi X-ray Timing Explorer (RXTE - see Greiner, Morgan, and Remillard
(1996) and references therein for further details regarding the
intrument and data modes).  We present the X-ray intensity for July
10, 1998 in Figure 1(b).

	The most obvious features in the IR lightcurve in Figure 1 are
6 faint flares.  The flares have peak amplitudes of $\sim 0.3-0.6$ mJy
(or $\sim 5-10$ mJy de-reddened for $A_K \sim 3$ mag) -- more than an
order of magnitude fainter than the Class B flares (i.e. Fender, et
al.  1997; Eikenberry et al., 1998a).  They have typical durations of
$\sim 500$ seconds, and are roughly symmetric in time.  Furthermore,
they repeat on timescales from $\sim 30-60$ minutes.  When
simultaneous X-ray coverage is available, the IR flares appear to be
associated with rapid X-ray fluctuations (Fig. 1b).  Inspection with
an expanded timescale shows several interesting aspects of these
pairings (Fig. 2).  The X-ray oscillations show a flare-dip-flare
morphology.  X-ray hardness ratios show that the dips are very soft
(see also Figure 4 d-f), as opposed to the hard X-ray dips associated
with Class B IR/radio flares.  Furthermore, the rises of the IR flares
in Figure 2 appear to {\it precede} the X-ray oscillations.  Note that
for the first 2 X-ray dips, there are IR flares $\sim 1500-1800$
seconds later, suggesting a possible correspondence between X-ray dips
and highly delayed IR flares.  However, if this were the case, we
would expect X-ray dips at $\sim 24600$s and $\sim 30300$s, to match
the observed IR flares at 26200s and 31900s.  Since we do not see
X-ray dips at these times, we conclude that the actual IR/X-ray
correspondence has IR flares preceding X-ray dips by $\sim 200-600$ s.
{\it Thus, these observations are the first to clearly demonstrate the
time ordering of associated X-ray dips and IR flares in GRS 1915+105.}

\subsection{August 1997 Observations}

	We also observed GRS 1915+105 simultaneously with the Palomar
5-m telescope and RXTE on 13-15 August 1997 (see also Eikenberry et
al., 1998a,b).  The basic obervational parameters were similar to
those for July 1998 described above.  On 14-15 August 1997, we
observed a series of Class B IR flares with their corresponding X-ray
cycles of hard dips and thermal flares.  We also noted that at times
the IR flux from GRS 1915+105 showed a noticeable quasi-steady IR
excess (Figure 3a), much lower than the flux levels from the Class B
flares themselves, but higher than the apparent baseline IR emission
of $\sim 3.6$ mJy on those nights.  Interestingly, the episodes of
excess IR emission appear to be associated with rapid X-ray
oscillations (Figure 3b) that seem to resemble the X-ray cycles seen
in July 1998 (Figure 2).  Motivated by the X-ray/IR association we
observed in the July 1998 data, we performed detailed X-ray spectral
analyses of X-ray oscillations in both epochs.  Figure 4 shows the
resulting best-fit parameters to typical X-ray oscillations from both
epochs at 1-second time resolution using the XSPEC package and an
absorbed multi-temperature blackbody + power-law model (identical to
those described in Muno et al., 1999).  Not only are the morphologies
of the events quite similar (although the August 1997 cycle is $\sim
3$ times {\it faster}), but the key spectral parameters of blackbody
temperature and power-law index seem to evolve in a virtually
identical manner for both epochs.  {\it These similarities in both
morphology and spectrum confirm that the X-ray cycles from July 1998
and August 1997 are indeed the same phenomenon.}  Furthermore, note
that the blackbody temperature drops and the power-law index rises
during the X-ray dip, both of which effects cause a softening of the
X-ray spectrum during the dip.  The X-ray dips associated with Class B
flares, on the other hand, show a decrease in the BB temperature and a
marked {\it decrease} in the power-law index, making them spectrally
hard.  Thus, the events we discuss here differ from those associated
with Class B flares.

	Based on these results, we then hypothesize that the IR excess
seen in 14-15 August 1997 during the X-ray oscillations may be due to
faint infrared flares such as those seen in Figures 1-2.  Since the
X-ray oscillations are separated by $~\sim 20-40$ seconds in August
1997 and the typical width of the faint IR flares is $\sim 500$
seconds, many flares will be superposed on one another to create the
appearance of a quasi-steady IR excess such as we observe.  If we
assume that each X-ray oscillation in Figure 3(b) has an associated IR
flare and we approximate that flare as a gaussian with $\sim 0.3$ mJy
amplitude and 160 seconds FWHM (consistent with the faintest July 1998
flares), we calculate a predicted IR excess of 1.3 mJy.  This value is
a close match to the actual observed excess of $\sim 1.0$ mJy we
observed (Figure 3).

\section{Discussion}

	Based on these observations, we surmise that we have found a
new type of IR flare associated with X-ray oscillations in GRS
1915+105.  These events differ significantly from the previously-known
Class B events in their IR brightness as well as the timescale,
morphology, and spectral characteristics of the X-ray oscillations.
In keeping with our proposed classification scheme for such flares --
Class A being major ejection events and Class B being the $\sim
100-200$ mJy (de-reddened) IR/radio flares associated with hard X-ray
dips -- we assign these faint IR flares associated with soft X-ray
dips the label ``Class C''.

	The July 1998 observations are useful not only in allowing us
to identify this new phenomenon, but also in allowing us to determine
the timing relationship between the X-ray and IR oscillations.
Previous observations of Class B events (e.g. Eikenberry et al, 1998a)
have been unable to unambiguously determine whether the IR/radio
flares come from an ejection at the beginning of the preceding hard
X-ray dip, at its end, or simultaneously with a soft X-ray ``spike''
seen during the dip.  Mirabel et al. (1998) suggest that the ejection
occurs at the time of the spike, based on timing/flux arguments and an
expanding plasmoid (van der Laan) model for their IR/radio data.
However, this model predicts an IR peak flux density $\sim 20$ times
higher than observed, and thus this issue remains unresolved for now.

	There are several physical phenomena which might produce the
Class C behavior, but our understanding may be helped by recently
published X-ray/radio observations of Feroci et al. (1999).  Using
BeppoSAX and the Ryle Telescope, they report an X-ray event very
similar in both flux and spectral evolution to those we report here.
Furthermore, they observed a $\sim 40 $ mJy radio flare which peaked
$\sim 1000$ seconds {\it after} the X-ray event.  If we assume that
this is a Class C event, and furthermore that it had an (unobserved)
IR flare similar in flux density and timing to those we observed, then
we must conclude that the flares have a flat peak flux density over
several decades of frequency ($F_{\nu} \propto \nu^{-0.15}$), with
longer wavelengths delayed compared to shorter wavelengths.  This
behavior closely resembles that of Class B flares (Mirabel et al.,
1998), and thus suggests that the Class C flares are also due
to synchrotron emission from an expanding plasma bubble.

	The fact that the IR flares precede the X-ray oscillations
suggests an ``outside-in'' model for these events.  In such a model, a
disturbance far from the black hole propagates inward, first creating
the synchrotron flare.  Then as the disturbance reaches the innermost
portion of the accretion disk, which produces the majority of the
thermal X-ray flux, it creates the X-ray flare-dip-flare cycle.
Several possibilities may explain these observations.  If we assume
that Class C events are due to ejection events which occur {\it
before} the inner disk is perturbed, we must conclude that the
innermost portion of the accretion disk is {\it not} the site of
origin for the ejections, contrary to what is generally believed for
microquasars (and other relativistic jet systems).  An alternative
interpretation may be that the IR/radio flare comes from a plasma
bubble created by a magnetic reconnection event in the accretion disk,
which would generate a disturbance in the accretion flow.  Theorists
have hypothesized that such reconnection events may be commonplace in
systems where jets are powered by magnetocentrifugal launching
mechanisms.  Yet another interpretation could be that the jets in GRS
1915+105 are not composed of discrete events, but are continuous
low-luminosity outflows punctuated by the appearance of occasional
high-luminosity shock events propagating through the flow (as has been
suggested for the case of relativistic jets in AGN).  In this case,
the Class C events could be due to a reverse shock propagating through
the jet back towards the disk.  As it nears the inner disk, the shock
would first produce a synchrotron bubble, generating the IR (and
eventually radio) flares, and then reach the inner disk itself to
disrupt the X-ray emission, as observed.

\section{Conclusions}

	We have reported a new type of IR/X-ray oscillation in the
microquasar GRS 1915+105.  These oscillations show faint ($\sim 0.5$
mJy) IR flares with durations of $\sim 500$ seconds, and are
associated with X-ray cycles of soft dips and thermal flares.  This
distinguishes them from previously known GRS 1915+105 behaviors which
show either major radio flares (Class A) or brighter ($\sim 100-200$
mJy) IR/radio flares accompanied by X-ray events with hard dips and
thermal flares (Class B).  Thus, we label these events as ``Class C''.

	Combining our observations with X-ray/radio observations of a
single Class C event by Feroci et al. (1999) indicates that the Class
C events are due to synchrotron emission from an expanding plasmoid.
Furthermore, in the Class C events the IR flare precedes the onset of
the X-ray cycle by several hundred seconds, suggesting an
``outside-in'' model for them.  Several possibilities exist for
explaining this behavior, including magnetic reconnection events in
the outer disk or reverse shocks propagating through a continuous jet
medium.

\acknowledgments The authors would like to thank the members of the
Rossi X-ray Timing Explorer team, without whose work none of these
investigations would have been possible.  SE thanks R. Lovelace,
M. Romanova, and R. Taam for helpful discussions of these
observations.  This work was supported in part by NASA grant NAG
5-7941.

\vfill \eject

\begin{figure}
\vspace*{100mm} 
\includegraphics{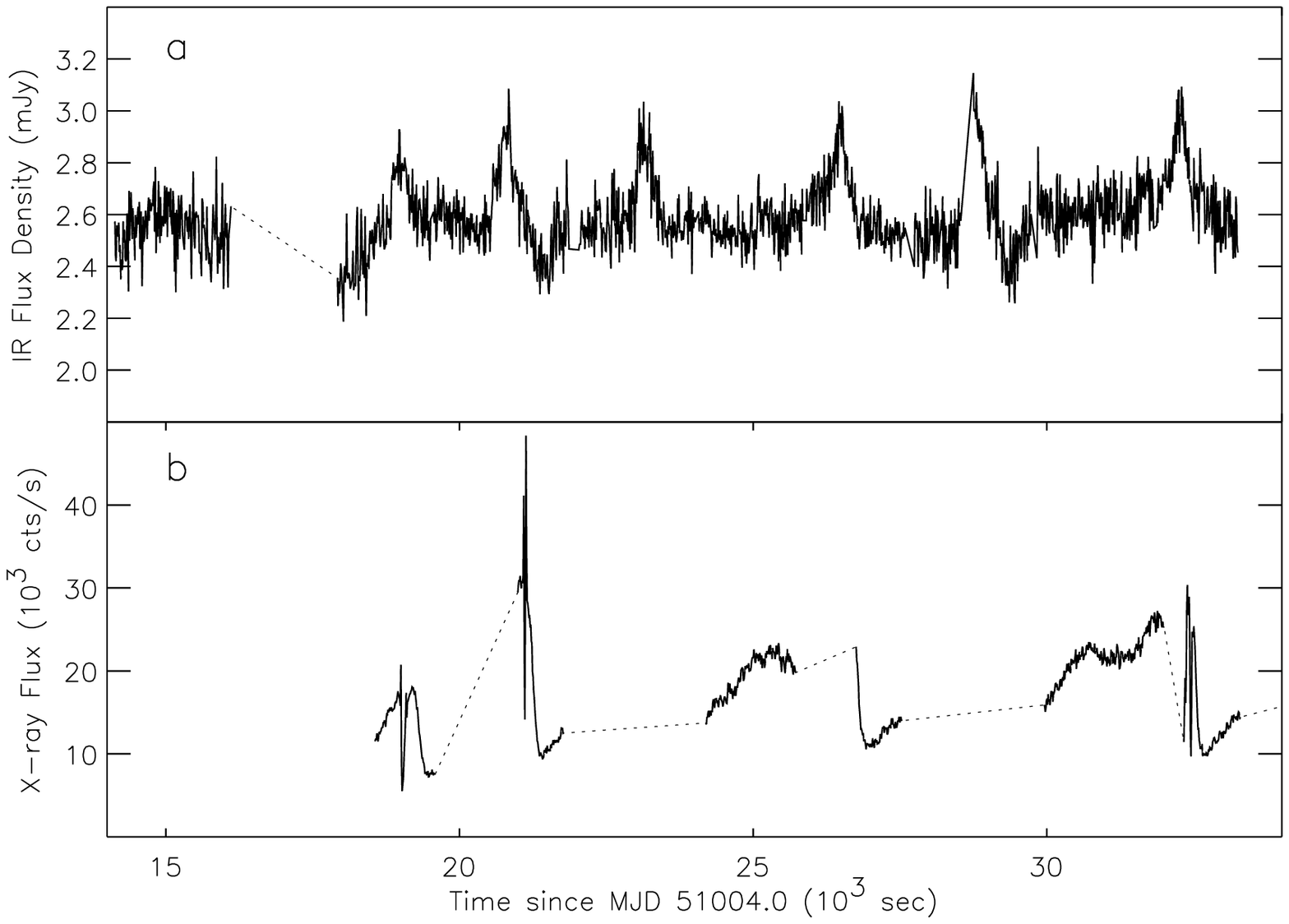}
\caption{\it Lightcurves for GRS 1915+105 on July 10, 1998 UTC.  Top
panel shows the IR ($2.2 \mu$m) flux density from the Palomar 5-m.
Bottom panel shows the PCA count rate from RXTE.  Both panels have
10-second time resolution.  Note that the IR flares are associated
with X-ray flare/dip cycles.}
\end{figure}

\begin{figure}
\vspace*{100mm}
\includegraphics{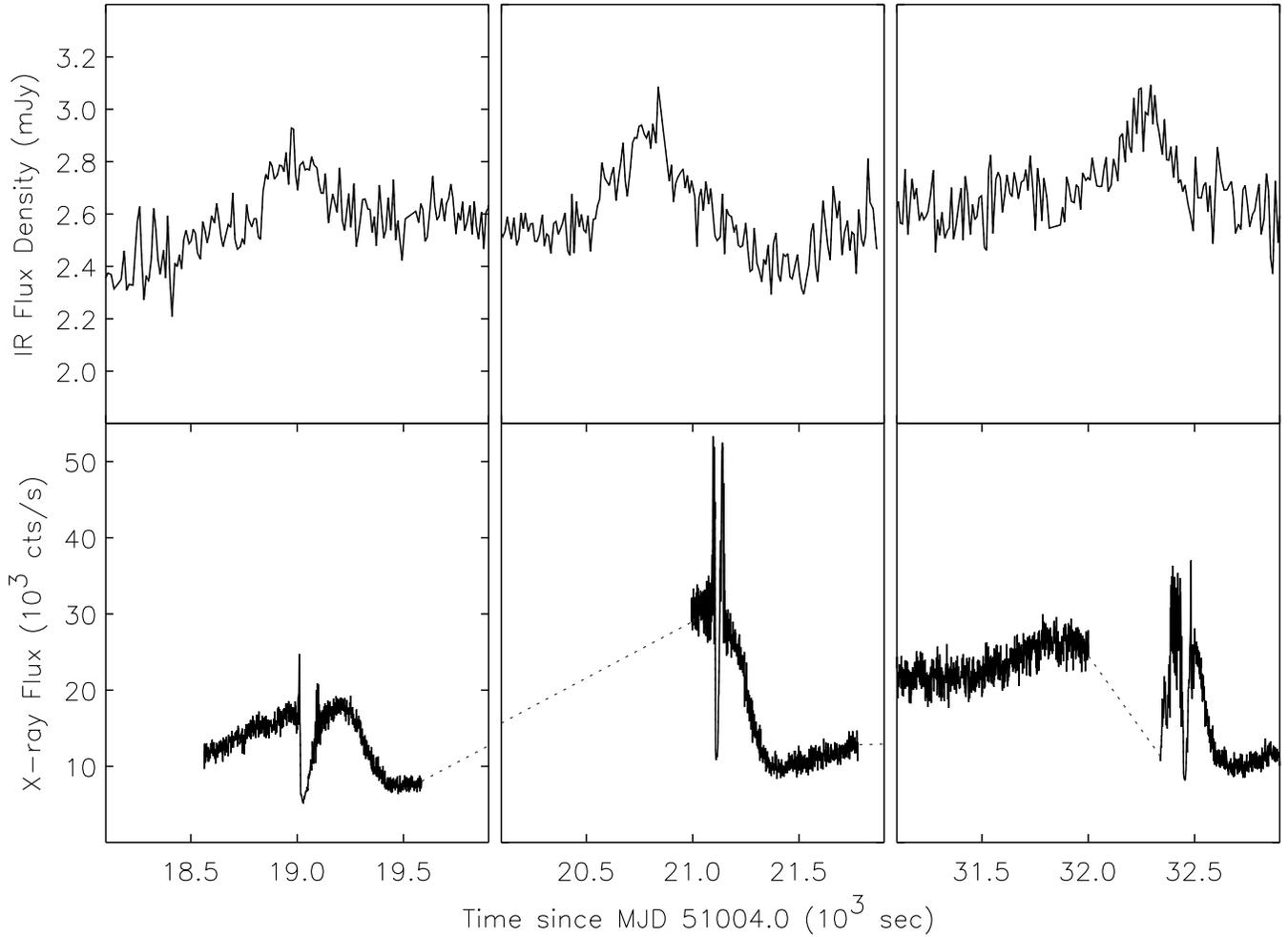}
\caption{\it Close-up views of simultaneous occurences of IR flares
with X-ray observations shown in Figure 1.  The top panels have
10-second time resolution, while the bottom panels have 1-second time
resolution (for improved signal-to-noise in the IR band).  Note that
the IR flares begin {\bf before} the X-ray flare-dip-flare pattern in
all 3 cases.}
\end{figure}

\begin{figure}
\vspace*{100mm}
\includegraphics{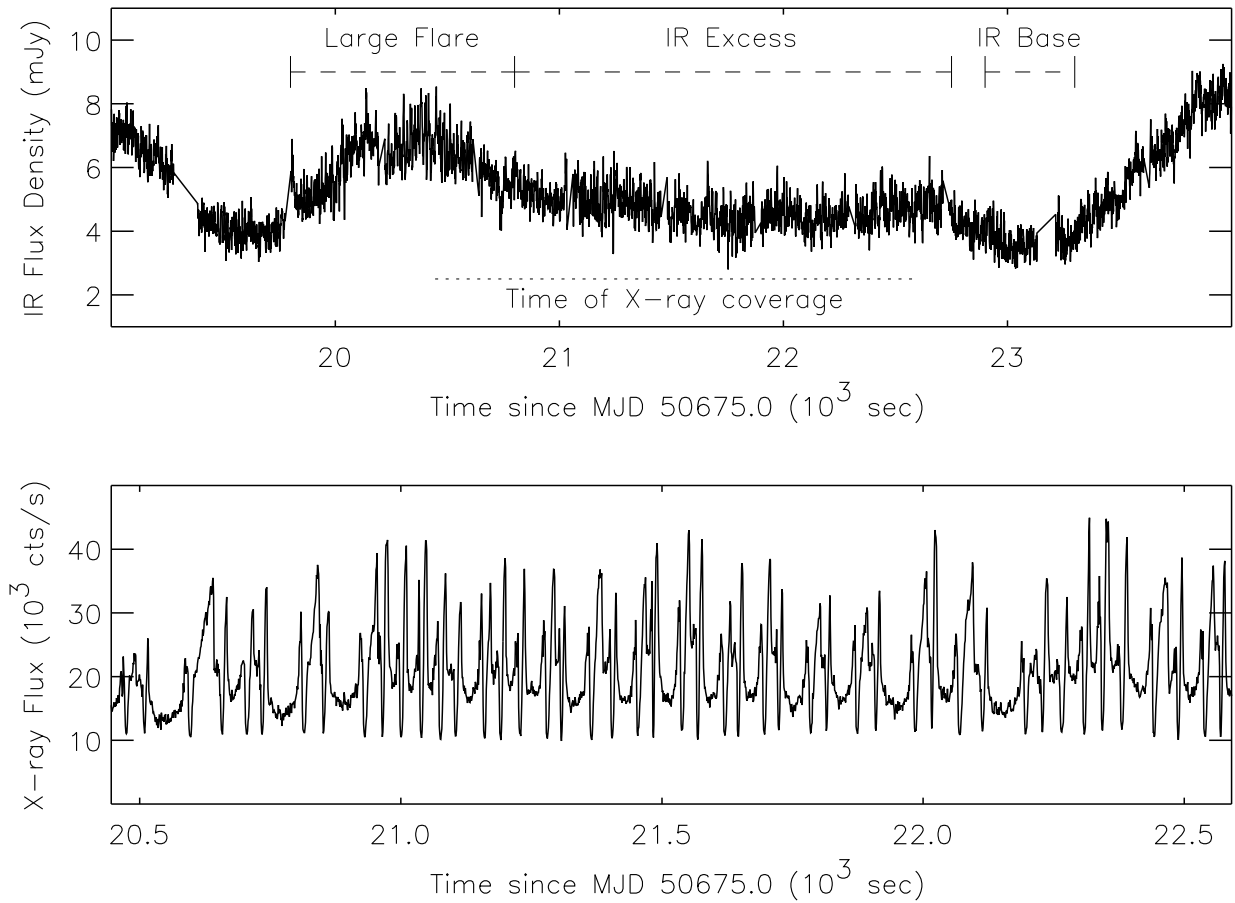}
\caption{\it Partial lightcurves for GRS 1915+105 on August 15, 1997
UTC (see also Eikenberry et al., 1998a).  The top panel shws the IR
($2.2 \mu$m) flux density from the Palomar 5-m telescope.  Features
labelled include a large flare associated with an X-ray
hard-dip/soft-flare, the quasi-steady IR excess associated with X-ray
soft-dip/soft-flare cycles, and the baseline IR emission.  The bottom
panel shows the X-ray flux from the PCA instrument on RXTE (time scale
has been expanded for clarity).  Note that the X-ray behavior
resembles that seen in Figure 2.}
\end{figure}

\begin{figure}
\vspace*{100mm}
\includegraphics{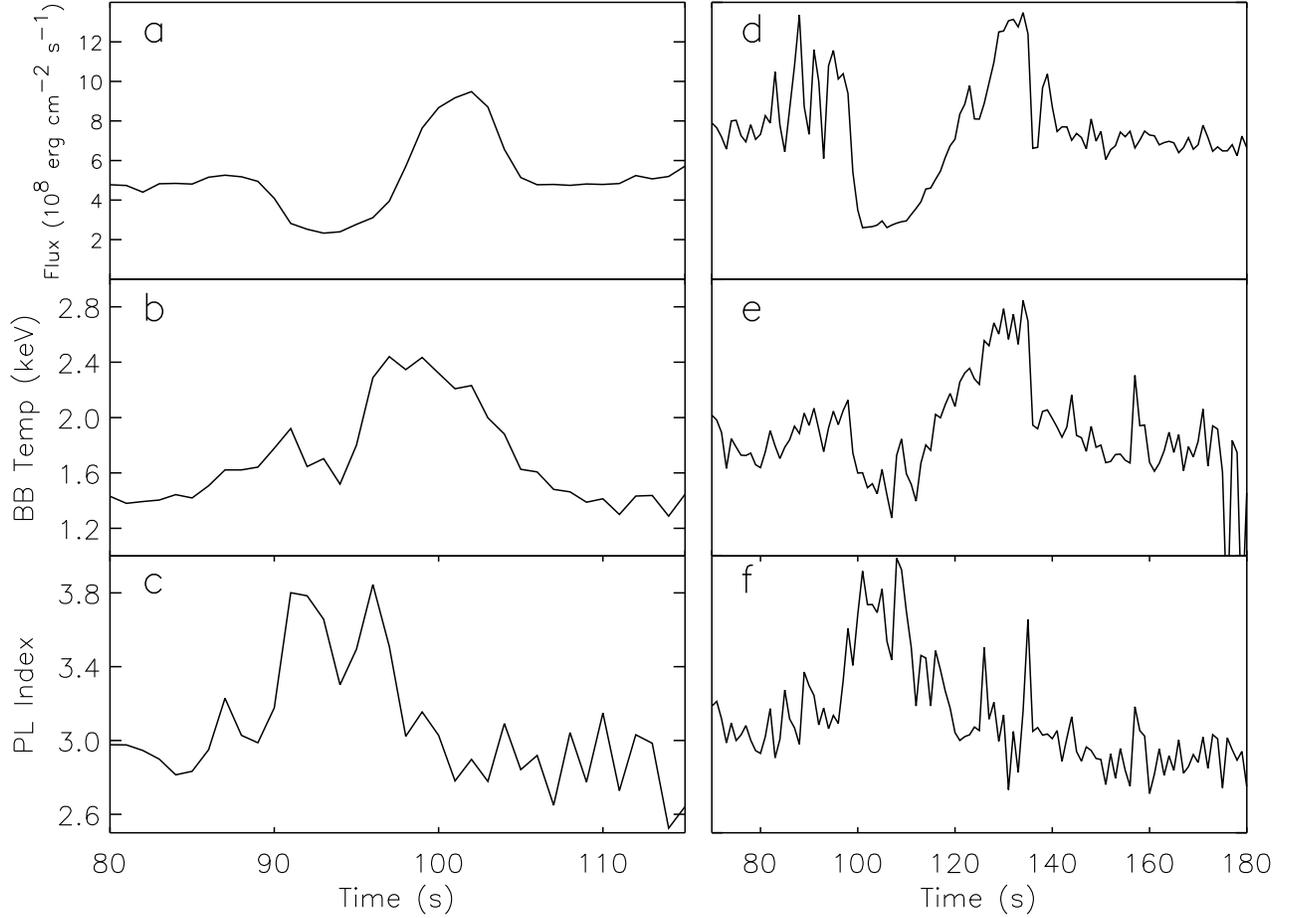}
\caption{\it X-ray spectral analyses of soft dip cycles from August
1997 (a-c -- see Figure 3) and July 1998 (d-f -- see Figure 2b) with
1-second time resolution.  The top row (a,d) shows the unabsorbed
X-ray flux in the 2-30 keV band.  The middle row (b,e) shows the
best-fit blackbody temperature, and the bottom row (c,f) shows the
best-fit power-law indices.  While the timescales differ by nearly a
factor of 3, the spectral characteristics of the events are nearly
identical.}
\end{figure}

\end{document}